\documentclass[12pt]{article}
\usepackage[utf8]{inputenc}
\usepackage{epsf,amsmath,amssymb,graphicx,dcolumn}
\usepackage{caption}
\usepackage{subcaption}
\usepackage{scalefnt,ulem,pstricks}
\usepackage{booktabs,multirow,tabularx}
\usepackage{cite}
\usepackage{cleveref}
\usepackage{color}
\usepackage{rotating}
\usepackage{microtype}
\usepackage{soul}
\definecolor{lightblue}{rgb}{.7,.8,1}

\renewcommand{\thefootnote}{\fnsymbol{footnote}}

\newcommand\nn{\nonumber}
\newcommand\f[2]{\frac{#1}{#2}}

\def\bal#1\eal{\begin{align}#1\end{align}}

\newcommand{\as}{\alpha_{\mathrm{S}}}

\newcommand{\mh}{m_{H}}

\newcommand{\pt}{\ensuremath{p_T}}

\newcommand{\citere}[1]{Ref.\cite{#1}}
\newcommand{\citeres}[1]{Refs.\cite{#1}}
\newcommand{\eqn}[1]{Eq.\,(\ref{#1})}
\newcommand{\neqn}[1]{Eqs.\,(\ref{#1})}
\newcommand{\fig}[1]{Fig.\,\ref{#1}}
\newcommand{\figs}[1]{Figs.\,\ref{#1}}

\newcommand{\muF}{\mu_{F}}
\newcommand{\muR}{\mu_{R}}

\newcommand{\msbar}{\overline{\text{MS}}}
\newcommand{\ep}{\epsilon}

\newcommand{\calO}{\mathcal{O}}

\usepackage{datetime}

\newdateformat{monthyeardate}{%
  \monthname[\THEMONTH] \THEYEAR}

\Crefname{figure}{Fig.}{Figs.}
\begin{document}
\begin{titlepage}
\renewcommand{\thefootnote}{\fnsymbol{footnote}}
\begin{flushright}
ZU-TH 43/16\\
PSI--PR--16-16\\
CERN-TH-2016-245
\end{flushright}
\vspace*{2cm}

\begin{center}
{\Large \bf Modeling BSM effects on the Higgs transverse-\\[0.4cm] momentum spectrum in an EFT approach}
\end{center}

\par \vspace{2mm}
\begin{center}
{\bf Massimiliano Grazzini$^{(a)}$},
{\bf Agnieszka Ilnicka$^{(a,b,c)}$},\\[0.2cm]
{\bf Michael Spira$^{(c)}$} and {\bf Marius Wiesemann$^{(a,d)}$}

\vspace{5mm}

$^{(a)}$ Physik-Institut, Universit\"at Z\"urich, 
CH-8057 Z\"urich, Switzerland 

$^{(b)}$ Physics Department, ETH Z\"urich, 
CH-8093 Z\"urich, Switzerland 

$^{(c)}$ Paul Scherrer Institute, CH-5232 Villigen PSI, Switzwerland

$^{(d)}$ CERN Theory Division, CH-1211, Geneva 23, Switzerland

\end{center}

\par \vspace{2mm}
\begin{center} {\large \bf Abstract} \end{center}
\begin{quote}
\pretolerance 10000

We consider the transverse-momentum distribution of a Higgs
boson produced through gluon fusion in hadron collisions.  At small
transverse momenta, 
the large logarithmic terms are resummed up to next-to-leading-logarithmic  (NLL)
accuracy.  The resummed computation is consistently matched to the next-to-leading-order (NLO)
result valid at large transverse momenta.
The ensuing Standard
Model prediction is supplemented by possible new-physics effects
parametrised through three dimension-six operators related to the
modification of the top and bottom Yukawa couplings, and to the
inclusion of a point-like Higgs-gluon coupling, respectively.  We
present resummed transverse-momentum spectra including the effect of
these operators at NLL+NLO accuracy and study their impact on
the shape of the distribution. We find that such modifications, while
affecting the total rate within the current uncertainties,
can lead to significant distortions of the spectrum. The proper
parametrization of such effects becomes increasingly important for
experimental analyses in Run II of the LHC.
    
\end{quote}

\vspace*{\fill}
\begin{flushleft}
\monthyeardate\today

\end{flushleft}
\end{titlepage}

\section{Introduction}

The discovery of a 125 GeV resonance \cite{ATLASdisc, CMSdisc}
that is compatible with the Standard-Model (SM) Higgs boson
\cite{Khachatryan:2016vau} dominated the recent years' activities in
particle physics. The existence of a Higgs boson is a basic prediction
of spontaneous symmetry breaking via a scalar sector \cite{Higgs:1964ia,
Higgs:1964pj, Englert:1964et, Guralnik:1964eu, Higgs:1966ev,
Kibble:1967sv}. The Higgs mechanism preserves the full gauge symmetry
and renormalizability of the SM \cite{'tHooft:1971rn, 'tHooft:1972fi}.
The most important Higgs-production channel at the LHC is gluon fusion,
which, despite being a loop-induced process, is highly enhanced by the
dominance of the gluon densities \cite{ggfus}.  The QCD corrections are
known up to N$^3$LO in the limit of heavy top quarks
\cite{Djouadi:1991tka, Dawson:1990zj, Dawson:1993qf, Harlander:2001is,
Harlander:2002wh, Anastasiou:2002yz, Ravindran:2003um, Marzani:2008az,
Gehrmann:2011aa, Anastasiou:2013srw, Anastasiou:2013mca,
Kilgore:2013gba, Li:2014afw, Anastasiou:2014lda, Anastasiou:2015ema,
Anastasiou:2016cez}, while the full quark-mass dependence is only known
up to NLO \cite{Graudenz:1992pv, Spira:1995rr, Harlander:2005rq,
Anastasiou:2009kn}. At NNLO subleading terms in the heavy top expansion
\cite{Harlander:2009bw, Pak:2009bx, Harlander:2009mq, Pak:2009dg} and
leading contributions to the top+bottom interference
\cite{Mueller:2015lrx} are known.
The limit of heavy-top quarks has also
been adopted for threshold-resummed calculations \cite{Kramer:1996iq,
Spira:1997dg, Catani:2003zt, Moch:2005ky, Ravindran:2005vv,
Ravindran:2006cg, Idilbi:2005ni, Ahrens:2008nc, deFlorian:2009hc,
deFlorian:2014vta, Catani:2014uta, Bonvini:2014tea}. The inclusion of
finite quark-mass effects in the resummation has been considered
recently \cite{deFlorian:2012yg, Bonvini:2014joa, Schmidt:2015cea}.

Kinematical distributions provide an important handle on the
determination of Higgs properties.  Among the most relevant observables
in this respect is the Higgs transverse-momentum ($p_T$) distribution.  First
results from the LHC Run I were presented by the ATLAS collaboration in
the $2\gamma$ and four-lepton final states \cite{atlas1,atlas2} and by
the CMS collaboration in the $2\gamma$ final state \cite{CMSpt}. In Run
2 these measurements can be extended to a larger range in the transverse
momentum with significantly higher statistics and accuracy after
accumulating up to ${\calO}(100~{\rm fb}^{-1})$ of luminosity. The
transverse-momentum spectrum provides more information
than the total cross section
and allows us to disentangle effects that remain hidden in the total
rates.  For example, it is the simplest measurement to shed light on the
nature of the Higgs coupling to gluons. The fact that the Higgs is a
scalar, gives an additional simplification in the modeling of the Higgs
$p_T$-spectrum, due to the factorization of production and decay in the
narrow-width approximation.

In the past years a significant amount of work has been done to improve
the theoretical predictions for the Higgs $p_T$ spectrum. The first
results at the lowest order (${\cal O}(\as^3)$) were known since long time
\cite{ptLO1,ptLO2} including the full quark-mass dependence. It took 
nearly ten years until the ${\cal O}(\as^4)$
corrections were computed \cite{ptNLO0,ptNLO1,ptNLO2,ptNLO3}.  These
were carried out in the heavy-top limit (HTL, i.e. $m_{t}^2 \gg M^2_H,
p^2_{TH}$). Finite top-mass effects on the Higgs $p_T$
distribution at ${\cal O}(\as^4)$ were estimated in
Refs.\,\cite{Harlander:2012hf,Neumann:2014nha,Neumann:2016dny}. 
Recently, results on Higgs+jet production at ${\cal O}(\as^5)$ were also obtained in the HTL
\cite{ptNNLO1,ptNNLO2,ptNNLO3}.

In the low-$p_T$ region ($p_T\ll \mh$), the convergence of the perturbative expansion
is spoiled by the presence of large logarithmic terms of the form $\as^n\ln^m (\mh^2/p_T^2)$.
In order to obtain reliable predictions also in this region, the large logarithmic terms must
be resummed to all orders \cite{Dokshitzer:1978hw,Parisi:1979se,Collins:1984kg,Bozzi:2005wk,Catani:2010pd}. 
It is then essential to consistently match the
resummed and fixed-order calculations in the intermediate \pt{} region, so as to obtain
accurate predictions in the entire region of transverse momenta.
In the case of the Higgs $p_T$ spectrum the resummation has been performed up to
next-to-next-to-leading logarithmic accuracy (NNLL) and matched
to the fixed-order NNLO result up to ${\cal O}(\as^4)$ in the HTL \cite{Bozzi:2005wk}. 
Finite quark-mass effects have been included in the resummed spectrum up to NLL+NLO \cite{Mantler:2012bj,Grazzini:2013mca}.
The recent computation of
the ${\cal O}(\as^5)$ corrections at high-$p_T$, together with new available
information on the logarithmic structure at the same order \cite{Li:2016ctv} would in principle allow
to extend the accuracy of this calculation.\footnote{Work in this direction has been recently presented in Ref.~\cite{Monni:2016ktx}.}

To consistently introduce deviations from the SM Higgs sector
explicit models beyond the SM (BSM) can be directly studied.
A complementary bottom-up framework is offered by
the {\it Standard Model Effective Field Theory} (SMEFT).
In such approach the Standard Model Lagrangian is extended by the inclusion of
operators of higher dimension (in first approximation: dimension six),
built from Standard Model fields and suppressed by powers of the New
Physics scale ($\Lambda$) \cite{Burges:1983zg, Leung:1984ni,
dim61,dim62}. This enables a theoretically consistent parametrisation of
model-independent effects of high-scale New Physics, which manifest themselves
through small deviations from the SM predictions. Many groups
translated the data already collected by the LHC as well as earlier
experiments into bounds on the Wilson coefficients of dimension-six SMEFT
operators (see e.g.,~\citeres{fits1,fits2,Ellis:2014jta,Dumont:2013wma,
Falkowski:2015fla, Butter:2016cvz}).

On the other
side, a significant effort has been devoted to supplement the tools used in the
modelling of LHC data with the effects of appropriate dimension-six
SMEFT operators (see e.g. \cite{Alloul:2013naa,Contino:2014aaa,Falkowski:2015wza,Zhang:2016snc,deAquino:2013uba,Artoisenet:2013puc}).
This is highly relevant since in this way
indirect BSM effects can be directly tested
in the experimental analyses.
The precision reached by the current
experiments will call for theoretical improvements and effects from
SMEFT operators beyond leading order
\cite{Passarino:2012cb}. Despite conflicting approaches followed in the literature,
SMEFT effects should be evaluated by including all possible operators
contributing to the observable (at a given order). Results for the total
Higgs production cross section including
modified top and bottom Yukawa
couplings and an additional direct $Hgg$ interaction
have been obtained at NNLO in Ref.~\cite{Brooijmans:2016vro} and at N$^3$LO
in Refs.~\cite{Harlander:2016hcx, Anastasiou:2016hlm}. Studies of the prospects of 
future LHC runs for the determination of Wilson coefficients 
were performed with the use of such tools in \citeres{Azatov:2016xik,ptdim63}.

The inclusion of dimension-six and dimension-eight operators in the
$p_T$-spectrum has been considered in
Refs.~\cite{ptdim61,ptdim62,ptdim63,Maltoni:2016yxb} and \cite{ptdim81,ptdim82},
respectively.  Strategies for extracting information on the Higgs-gluon
couplings from the measurements were studied in Ref.~\cite{ptdim63}.
Most of the above studies, however, are limited to the high-$p_T$ region
of the spectrum, and do not include small-$p_T$ resummation. In this
paper we present a computation of the resummed $p_T$-spectrum at
NLL+NLO accuracy, with the inclusion of a set of dimension-six
operators relevant for Higgs boson production. 

The paper is organised as follows. In Sec.~2 we discuss the effects on the Higgs production cross section from the inclusion of the dimension-six operators and we explicitly evaluate the modifications of the inclusive LO cross section. In Sec.~3 we 
outline the computation of the $p_T$ spectrum of the Higgs boson, review
the formalism of transverse-momentum resummation required at small $p_T$ and
describe our NLL+NLO calculation. In Sec.~4 we present our results for the $p_T$ spectrum
and study its sensitivity to BSM effects of the dimension-six operators.
In Sec.~5 we summarize our results and provide some concluding remarks.

\section{Effective operators and their impact on the Higgs production cross section}

\begin{figure}[tp]
\begin{center}
\includegraphics[width=.8\textwidth]{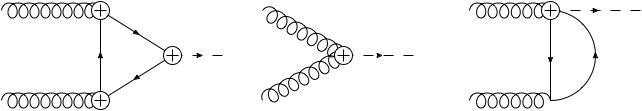}\vspace{0.2cm}
\caption[]{\label{fig:Fgr}{Feynman diagrams contributing to $gg\rightarrow H$ production at LO. The possible insertions of dimension-six operators are marked by a cross in a circle.}}
\end{center}
\end{figure}

We consider the effective Lagrangian
\begin{equation}
{\cal L}={\cal L}_{SM}+\sum_i \f{c_i}{\Lambda^2}{\cal O}_i
\end{equation}
where the SM is supplemented by the inclusion of a set of dimension-six
operators describing new physics effects at a scale $\Lambda$ well above
the electroweak scale.  In our study we consider the following four
operators
\begin{align}
{\cal O}_1 &= |H|^2 G^a_{\mu\nu}G^{a,\mu\nu}\,,\quad {\cal O}_2 = |H|^2 \bar{Q}_L H^c u_R + h.c.\,,\\ {\cal O}_3 &= |H|^2 \bar{Q}_L H d_R + h.c.\,,\quad {\cal O}_4 = \bar{Q}_L H \sigma^{\mu\nu}T^a u_R G_{\mu\nu}^a + h.c.
\end{align}
These operators, in the case of single Higgs production, may be expanded as:
\begin{align}
\label{eq:o1}
\f{c_1}{\Lambda^2}\,{\cal O}_1 &\rightarrow \frac{\as}{\pi v} c_g h  G^a_{\mu\nu}G^{a,\mu\nu}\,,\\
\label{eq:o2}
\f{c_2}{\Lambda^2}\,{\cal O}_2 &\rightarrow \frac{m_t}{v} c_{t} h \bar{t} t\,,\\
\label{eq:o3}
\f{c_3}{\Lambda^2}\,{\cal O}_3 &\rightarrow \frac{m_b}{v} c_{b} h \bar{b} b\,,\\
\label{eq:o4}
\f{c_4}{\Lambda^2}\,{\cal O}_4 &\rightarrow c_{tg}\f{g_S m_t}{2v^3} (v+h)G_{\mu\nu}^a({\bar t}_L\sigma^{\mu\nu}T^a t_R+h.c)\,.
\end{align}

The operator ${\cal O}_1$ corresponds to a contact interaction between
the Higgs boson and gluons with the same structure as in the heavy-top
limit of the SM. The operators ${\cal O}_2$ and  ${\cal O}_3$ describe
modifications of the top and bottom Yukawa couplings.  The operator
${\cal O}_4$ is the chromomagnetic dipole-moment operator, which
modifies the interactions between the gluons and the top
quark\footnote{In this analysis we do not consider the contribution of
the chromomagnetic dipole operator of the bottom quark.} (here
$\sigma^{\mu\nu}=\f{i}{2}\left[\gamma^\mu,\gamma^\nu\right]$).  In our
convention, based on the SILH basis \cite{SILH0,SILH}, we express the
Wilson coefficients as factors in the canonically normalized Lagrangian. 

The coefficients $c_t, c_b$ and $c_g$ can be probed in Higgs boson processes.
In particular, $c_t$ (and $c_b$) may be measured in the $t\bar{t}H$ (and
$b\bar{b}H$) production modes.\footnote{See Refs.\,\cite{Beenakker:2001rj,
Reina:2001sf, Beenakker:2002nc,Dawson:2002tg} and
Refs.\,\cite{Dittmaier:2003ej, Dawson:2003kb, Harlander:2011fx, Harlander:2010cz,Wiesemann:2014ioa,
Harlander:2014hya}, respectively, and references therein.} The
coefficient $c_b$ can also be accessed through the decay $H \rightarrow
b\bar{b}$. The coefficient $c_{tg}$, instead, is constrained by top pair
production \cite{Franzosi:2015osa}.

We now consider the contribution of the effective operators in
\neqn{eq:o1}, \eqref{eq:o2} and \eqref{eq:o4} on the production cross section, while
omitting, for simplicity, the bottom contribution in \eqn{eq:o3}. The relevant Feynman 
diagrams are displayed in \fig{fig:Fgr}. The corresponding amplitude can be cast into the form
\begin{equation}
{\cal M}\left(g(p_1)+g(p_2)\to H\right)=i\f{\as}{3\pi v}\epsilon_{1\mu}\epsilon_{2\nu}
\left[p_1^\nu p_2^\mu-(p_1p_2)g^{\mu\nu}\right]F(\tau)\, ,
\end{equation}
where $\tau=4m_t^2/\mh^2$ and $\epsilon_1$ and $\epsilon_2$ are the
polarization vectors of the incoming gluons.  The contribution of the
chromomagnetic operator to the
function $F(\tau)$ has been addressed in the literature with
contradicting results \cite{Choudhury:2012np,Degrande:2012gr} (see also
Ref.~\cite{Chien:2015xha}).  In Ref.~\cite{Choudhury:2012np} it is found
that the UV divergences in the bubble and triangle contributions cancel
out. In the revised version of Ref.~\cite{Degrande:2012gr} it is instead
stated that the UV divergence is present, and it has to be reabsorbed
into the coefficient $c_g$.

Our results are consistent with the latter statement. We find
\begin{equation}
F(\tau)=\Gamma(1+\ep)\left(\f{4\pi\mu^2}{m_t^2}\right)^\ep \left(c_t F_1(\tau)+c_{g0} F_2(\tau)+Re(c_{tg})\f{m_t^2}{v^2} F_{30}(\tau)\right)\, ,
\end{equation}
where
\begin{align}
F_1(\tau)&=\f{3}{2}\tau \left[1+(1-\tau)f(\tau)\right]\,,\\
F_2(\tau)&=12\, ,\\
F_{30}(\tau)&=\f{6}{\ep}+3\left[1-\tau f(\tau)-2g(\tau)\right]\, ,
\end{align}
with the functions
\begin{align}
f(\tau) & = \left\{ \begin{array}{ll}
\displaystyle \arcsin^2 \frac{1}{\sqrt{\tau}}  & \tau \ge 1 \\
\displaystyle - \frac{1}{4} \left[ \ln \frac{1+\sqrt{1-\tau}}
{1-\sqrt{1-\tau}} - i\pi \right]^2  & \tau < 1
\end{array} \right. \;.
\end{align}
\begin{align}
g(\tau) & = \left\{ \begin{array}{ll}
\displaystyle \sqrt{\tau-1}\arcsin \frac{1}{\sqrt{\tau}}  & \tau \ge 1 \\
\displaystyle \sqrt{1-\tau}\left[ \ln \frac{1+\sqrt{1-\tau}}
{1-\sqrt{1-\tau}} - i\pi \right]  & \tau < 1
\end{array} \right. \;.
\end{align}
The $1/\ep$ divergence can be reabsorbed in the $\overline{\rm MS}$
renormalization of the coefficient $c_g$:
\begin{equation}
c_{g0}=c_g(\muR)+\delta c_g
\end{equation}
with
\begin{equation}
\delta c_g=\f{m_t^2}{2v^2} Re(c_{tg})\Gamma(1+\ep)(4\pi)^\ep \left(-\f{1}{\ep}-\ln\f{\mu^2}{\muR^2}\right)\, ,
\end{equation}
where $\mu_R$ denotes the renormalization scale of $c_g$.
The final result reads
\begin{equation}
F(\tau)=c_t F_1(\tau)+c_{g}(\muR) F_2(\tau)+Re(c_{tg})\f{m_t^2}{v^2} F_{3}(\tau)\, ,
\end{equation}
where
\begin{equation}
F_3(\tau)=3\left(1-\tau f(\tau)-2g(\tau)+2\ln \f{\muR^2}{m_t^2}\right)\, .
\end{equation}
In the HTL $m_t^2\gg \mh^2$ we have
\begin{equation}
F_1(\tau)\to 1\, ,~~~~~~~~~
F_2(\tau)\to 12\, ,~~~~~~~~
F_3(\tau)\to 6\left(\ln \f{\muR^2}{m_t^2}-1\right)\, .\nn
\end{equation}
In the SM we have $c_t=1$ and $c_g=c_{tg}=0$, so that $F(\tau)\to F_1(\tau)$.
In Ref.~\cite{Franzosi:2015osa} data on top production are used to
extract constraints on $c_{tg}$. The resulting region of allowed values of $c_{tg}$ has been
found to be
\begin{equation}
-0.04 \lesssim c_{tg} \lesssim 0.04\, .
\end{equation}
The impact on the total cross section is less than 20\%.
We conclude that, although smaller than the impact of $c_g$, the effect
of $c_{tg}$ can still be important.  We note, however, that the
chromomagnetic operator provides a contribution which is formally ${\cal
O}(\lambda_t^2)$ with respect to the others. In a strict expansion in
$\as$ it can be neglected. This is what we will do in the next Section.
Focusing on the impact of $c_t$ and $c_g$, we  note that the total cross
section alone does not allow us to disentangle the coefficients $c_g$
and $c_t$: \begin{align} \label{sigmacgct} \sigma \approx |12 c_g +
c_{t}|^2 \sigma_{SM}\ \ (HTL)\,.  \end{align} As already noted in the
literature \cite{ptdim62}, the transverse momentum spectrum allows us to
break this degeneracy.

\section{Transverse-momentum spectrum}

We consider the inclusive hard-scattering process
\begin{equation}
h_1(p_1)+h_2(p_2)\to H(p_T)+X
\end{equation}
where the colliding hadrons $h_1$ and $h_2$ with momenta $p_1$ and $p_2$
produce the Higgs boson $H$ with transverse momentum $p_T$
accompanied by an arbitrary and undetected
final state $X$.  According to the QCD factorization theorem the
transverse-momentum cross section is evaluated as
\begin{equation}
\f{d\sigma}{dp_T^2}(p_T,s)=\sum_{a_1a_2}\int_0^1dx_1dx_2 f_{a_1/h_1}(x_1,\muF^2)f_{a_2/h_2}(x_2,\muF^2)
\f{d{\hat \sigma}_{H,a_1a_2}}{dp_T^2}(p_T,{\hat s},\as(\muR^2),\muR^2,\muF^2)\, ,
\end{equation}
where $f_{a/h}(x,\muF^2)$ are the parton densities of the colliding
hadrons at the factorization scale $\muF^2$, $d{{\hat \sigma}_{H,a_1a_2}}/dp_T^2$
is the partonic cross section, ${\hat s}=x_1x_2 s$ is the partonic
centre-of-mass energy, and $\muR$ is the renormalization
scale\footnote{Throughout the paper we use parton densities as defined
in the $\msbar$ factorization scheme and $\as(q^2)$ is the QCD running
coupling in the $\msbar$ renormalization scheme.}.  In the low-$p_T$
region ($p_T \ll \mh$), the perturbative expansion is affected by large
logarithmic terms of the form $\as^n\ln^m(\mh^2/p_T^2)$, with $1\le m\le
2n$. This results in a singular behaviour of the cross section as
$p_T\rightarrow 0$. To cure this problem we need to resum these terms to
all orders in $\as$. To properly account for transverse-momentum
conservation, the resummation is carried out in impact parameter ($b$)
space \cite{Curci:1979bg,Parisi:1979se,Collins:1984kg}. In this paper we
use the formalism of Ref.~\cite{Bozzi:2005wk}. The partonic
transverse-momentum cross section is decomposed as
\begin{align}
\label{eq:resfin}
\frac{d{\hat \sigma}_{H,ab}}{d p_T^2}=
\frac{d{\hat \sigma}_{H,ab}^{\rm (res.)}}{d p_T^2}+\left(
\frac{d{\hat \sigma}_{H,ab}}{d p_T^2}
-\frac{d{\hat \sigma}_{H,ab}^{\rm (res.)}}{d p_T^2}
\right)_{\rm f.o.}\, .
\end{align}
The first term, $d\sigma_{H,ab}^{\rm (res.)}$, on the right-hand side of
Eq.~(\ref{eq:resfin}) contains all the logarithmically enhanced terms,
and is evaluated by resumming them to all orders.  The second term is
finite, and can be computed by fixed-order truncation of the
perturbative series: It is obtained by computing the standard
fixed-order result valid at large $p_T$ and subtracting the expansion of
the resummed term at the same fixed order.  This matching procedure
ensures that the resummed and fixed-order components are combined to
achieve uniform formal accuracy from the small- to the large-$p_T$
region. 

The explicit expression of the resummed component is
\begin{equation}
\frac{d{\hat \sigma}_{H,a_1a_2}^{\rm (res.)}}{d p_T^2}=\frac{\mh^2}{{\hat s}}
\int_0^\infty db\, \frac{b}{2}J_0(bp_T){\cal W}_{a_1a_2}(b,\mh,{\hat s};\as)\, ,
\end{equation}
where $J_0(x)$ is the $0$th-order Bessel function and the factor ${\cal W}$ embodies
the all-order resummation of the large logarithmic terms.
The all-order structure of ${\cal W}_{a_1a_2}$ is better expressed by considering the $N$-moments with respect to $z=\mh^2/{\hat s}$ at fixed $\mh$ and is given by
\begin{equation}
\label{eq:W}
{\cal W}_N(b,\mh,\as)={\cal H}_N(\mh,\as,\mh^2/Q^2)\exp\{{\cal G}_N(\as,{\tilde L},\mh^2/Q^2)\}
\end{equation}
where
\begin{equation}
\label{eq:Ltilde}
{\tilde L}=\ln\left(Q^2b^2/b_0^2+1\right)
\end{equation}
and $b_0=2e^{-\gamma_E}$ ($\gamma_E=0.5772...$ is the Euler number).
The function ${\cal H}_N$ in Eq.~(\ref{eq:W}) does not depend on the impact parameter $b$ and can thus
be expanded in powers of $\as$ as
\begin{equation}
{\cal H}_N(\mh,\as,\mh^2/Q^2)=\sigma_0(\as,\mh)\Bigg[1+\left(\frac{\as}{\pi}\right){\cal H}^{(1)}_N+\left(\frac{\as}{\pi}\right){\cal H}^{(2)}_N+\dots\Bigg]
\end{equation}
where $\sigma_0(\as,\mh)$ is the lowest order partonic cross section.
The dependence on $b$ is fully contained in the exponent ${\cal G}_N(\as,{\tilde L},\mh^2/Q^2)$ whose expansion reads
\begin{equation}
{\cal G}_N(\as,{\tilde L},\mh^2/Q^2)={\tilde L}g^{(1)}(\as,{\tilde L})+g^{(2)}_N(\as,{\tilde L})+\f{\as}{\pi}g^{(3)}_N(\as,{\tilde L})+\dots
\end{equation}
where $g^{(1)}$ controls the leading logarithmic (LL) terms, $g^{(2)}_N$ the NLL terms, and so forth.
The formalism we have briefly recalled defines a systematic expansion of Eq.~(\ref{eq:resfin}), whose terms are denoted
by NLL+NLO, NNLL+NNLO and so forth. The first label in this notation denotes the logarithmic accuracy, while the second one is referred to the accuracy of the fixed-order calculation.
In particular, the NLL+NLO accuracy is obtained by computing the resummed component including
the coefficient ${\cal H}^{(1)}$ together with the functions $g^{(1)}$ and $g^{(2)}$, and by matching it to the ${\cal O}(\as^3)$ fixed-order result valid at high $p_T$.
The NNLL+NNLO accuracy is obtained by including also the coefficient ${\cal H}^{(2)}$ and the function $g^{(3)}_N$,
and the finite component to ${\cal O}(\as^4)$.

The scale $Q$ appearing in Eq.~(\ref{eq:W}), called resummation scale, parametrizes the arbitrariness
in the resummation procedure. Its role is analogous to the role played by the renormalization (factorization) scale
in the renormalization (factorization) procedure. 
Although ${\cal W}_N$ does not depend on $Q$ when evaluated at all perturbative orders,
an explicit dependence on $Q$ appears when the logarithmic expansion is truncated at a given order.
In particular, since the scale $Q$ appears in the definition of the large logarithmic term ${\tilde L}$,
the resummation scale sets the scale up to which the resummation is effective.

As is well known (see Sect. 3 in Ref.~\cite{Bozzi:2005wk}),
the extrapolation of the resummed result at large
transverse momenta, where the resummation cannot improve the accuracy of
the fixed-order expansion, may lead to unjustified large uncertainties
and ensuing lack of predictivity.  In the numerical implementation of
Eq.~(\ref{eq:resfin}) we thus apply a smooth switching procedure as
described in Ref.~\cite{deFlorian:2012mx}(see in particular
Eqs.~(13)-(15))\footnote{In the present paper the switching parameters
are chosen as $p_T^{sw} = 125$ GeV and  $\Delta p_T = 75$ GeV.}.  We
also point out that, due to the definition of the logarithmic parameter in Eq.~(\ref{eq:Ltilde}),
the formalism of Ref.~\cite{Bozzi:2005wk} fulfils a {\it unitarity constraint} such that, upon integration
over $p_T$, the customary fixed-order prediction for the inclusive cross section is recovered.
More precisely, by performing the
resummation at NLL accuracy and including the fixed-order result up to
$O(\as^3)$ we obtain NLL+NLO accuracy, and the integral of the spectrum
is fixed to the NLO total cross section. Despite the switching procedure
discussed above the $p_T$ spectra we are going to present fulfil such
unitary constraint to better than 1\%.

Top- and bottom-mass effects can be included in the resummed spectrum
along the lines of
Refs.~\cite{Mantler:2012bj,Grazzini:2013mca}.\footnote{For studies of the
resummed $p_T$ spectrum in explicit BSM models see for example
Refs.~\cite{Bagnaschi:2011tu,Harlander:2014uea,Mantler:2015vba,Liebler:2016dpn}.}  The
inclusion of the top mass does not lead to complications: since $m_t\sim
\mh$ the computation of the $p_T$ spectrum is still a {\it two scale}
problem. The inclusion of the bottom-mass instead is more difficult.
Since $m_b\ll \mh$, the computation of the $p_T$ spectrum becomes a {\it
three scale problem}, whose all-order solution is far from being
trivial.\footnote{For a recent contribution on this subject see
Ref.~\cite{Melnikov:2016emg}.} In Ref.~\cite{Grazzini:2013mca} a simple
solution to this problem was proposed, which implies a choice of
different resummation scales for the top and bottom contributions. In
particular, since, as discussed above, the resummation scale is the
scale up to which the resummation is effective, it was suggested to
choose for the bottom contribution a {\it lower} scale as compared to
the top contribution. In Ref.~\cite{Harlander:2014uea} this approach has
been extended to consider three different resummation scales for the top
contribution, the bottom contribution, and the top-bottom interference.
We will follow such approach in the next Section.

We now discuss the inclusion of BSM effects in the computation. The operators in Eqs.~(\ref{eq:o1})-(\ref{eq:o3})
modify the computation of both the resummed and the fixed-order component in Eq.~(\ref{eq:resfin}).
However, by limiting ourselves to NLL+NLO, the computation can be greatly simplified.
Indeed, the fixed-order result valid at high $p_T$ can be obtained by introducing the $c_t$ and $c_b$ coefficients in the SM amplitude, and supplementing it with an additional contribution, proportional to $c_g$, which corresponds to the QCD amplitude computed in the HTL.
As for the resummed component, due to the universality of our formalism \cite{Catani:2013tia},
its only process dependent contribution is encoded in the coefficients $\sigma_0$ and ${\cal H}^{(1)}$, which are determined 
by the Born-level and one-loop amplitudes, respectively.
These amplitudes can also be easily obtained from the SM result by introducing the factors $c_t$ and $c_b$ where appropriate, and adding the point-like HTL amplitude in the SM with a coefficient $c_g$. We emphasize that the first EFT correction to the SM 
is obtained by interfering the EFT amplitudes with the corresponding SM contributions. With this strategy, we can obtain NLL+NLO predictions for the $p_T$ spectrum consistently including the effects of the dimension-six operators. Thanks to the unitarity constraint,
the integral of the $p_T$ spectrum exactly reproduces the fixed-order
NLO result obtained with the inclusion of the same operators. 

\section{Results}

In this Section we present our numerical results for the
transverse-momentum spectrum including the effect of dimension-six
operators.  Our implementation is based on the program {\sc HqT}
\cite{hqt1,hqt2}: a public tool for the computation of the analytic
transverse-momentum spectrum of the Higgs boson. The contributions from
finite top and bottom masses as well as the dimension-six operators are
consistently included up to NLL+NLO accuracy.  The fixed-order cross section is 
then cross checked with {\sc HIGLU} \cite{higlu} and {\sc HNNLO}
\cite{Catani:2007vq,Grazzini:2008tf,Grazzini:2013mca}.

We consider $pp$ collisions at $\sqrt{s}=13$ TeV. Our computation is
performed in the five-flavor scheme with the corresponding NLO set of the PDF4LHC2015
\cite{Butterworth:2015oua,Ball:2014uwa,Dulat:2015mca,Harland-Lang:2014zoa,Gao:2013bia,Carrazza:2015aoa}
parton distribution functions~(PDFs) and the respective value of the strong coupling constant. 
For the top and bottom quarks running in the loop and for their Yukawa couplings
on-shell masses $m_b = 4.92$\,GeV and $m_t = 172.5$\,GeV are used.  As discussed in
the previous Section, our computation of the NLL+NLO $p_T$ spectrum
fulfils a unitarity constraint, such that by integrating over $p_T$ we
recover the fixed-order NLO cross section. The appropriate scale choice
for such a resummed computation is of the order of the Higgs boson mass
$m_H$. In our study, however, we are also
interested in the high-$p_T$ region, where a dynamical scale choice has to be preferred. 
We thus proceed as follows: in the fixed-order
computation, which is valid at high $p_T$, our central renormalization
and factorization scales are set to
$\muR=\muF=\mu_0=\sqrt{\pt^2+\mh^2}/2$. In the resummed computation (and
its fixed-order expansion) we fix the central scales to
$\muR=\muF=\mh/2$.  To ensure a proper assignment of the resummation
scales for the individual contributions to the cross section we follow
the splitting of the SM cross section into a top, a bottom and an
interference contribution as suggested in \citere{Harlander:2014uea} and
assign different scales to each of these contributions. In particular we
choose:
\begin{align}
Q_t = \mh/2=62.5\,{\rm GeV}\,,\quad Q_b=4\,m_b=19.68\,{\rm GeV}\,,\quad Q_{\rm int}=\sqrt{Q_t\,Q_b}=35.07\,{\rm GeV}\,.
\end{align}

These values are justified by the findings of \citeres{Grazzini:2013mca,
Harlander:2014uea, Bagnaschi:2015bop, Bagnaschi:2015qta}.  As the top
contribution is essentially insensitive to the top-quark mass in the
small-\pt{} region, where resummation is relevant, we assign $Q_t$
also to the contribution with the point-like $ggH$ coupling, when
choosing $c_g\neq 0$.  In fact, regarding the splitting of the cross
section into the three contributions outlined above we consider the $c_g$ 
amplitude as part of the top amplitude.

\begin{figure}
\begin{center}
\begin{tabular}{cc}
\hspace*{-0.17cm}
\includegraphics[trim = 7mm -7mm 0mm 0mm, width=.36\textheight]{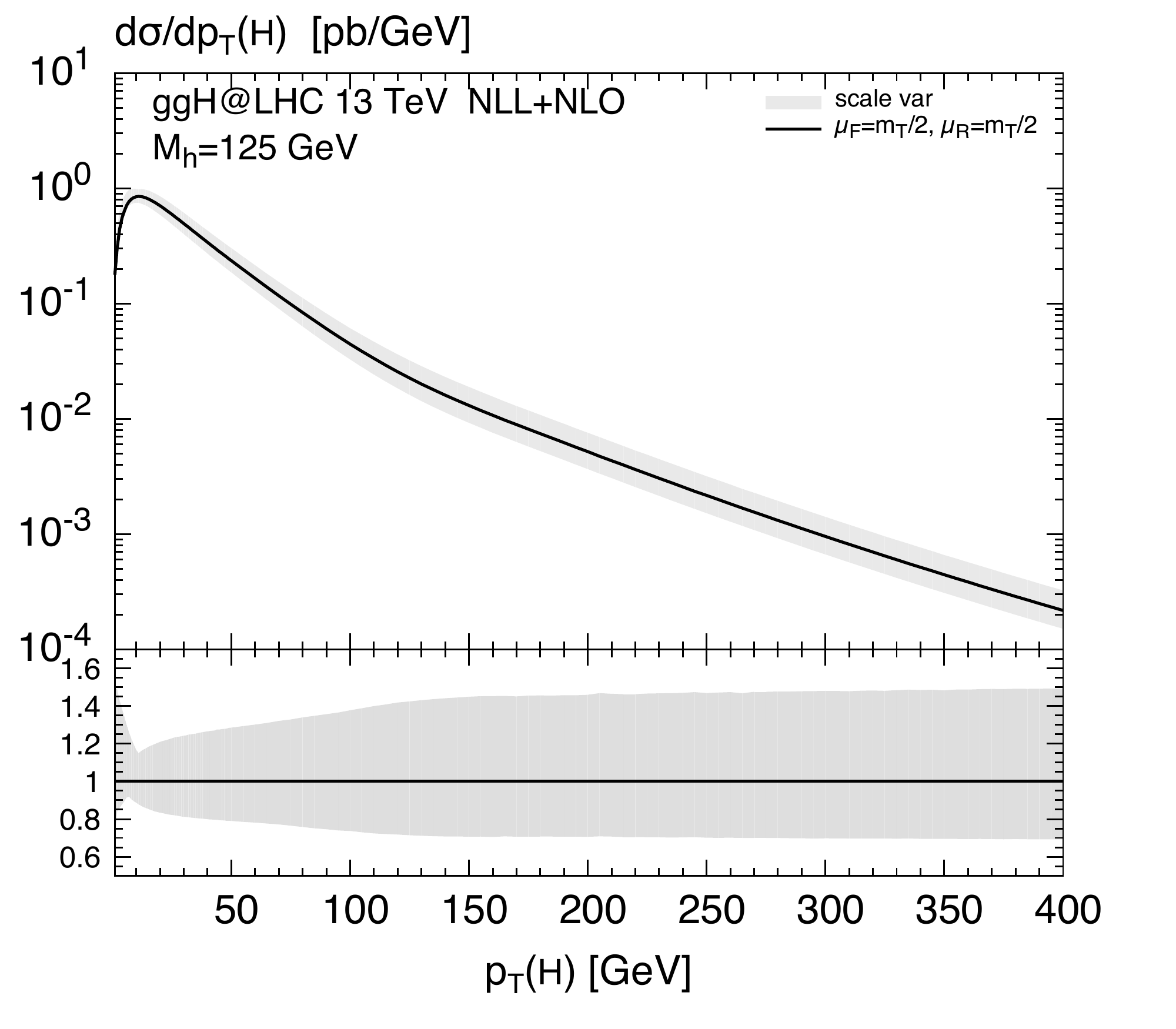} &
\includegraphics[trim = 7mm -7mm 0mm 0mm, width=.36\textheight]{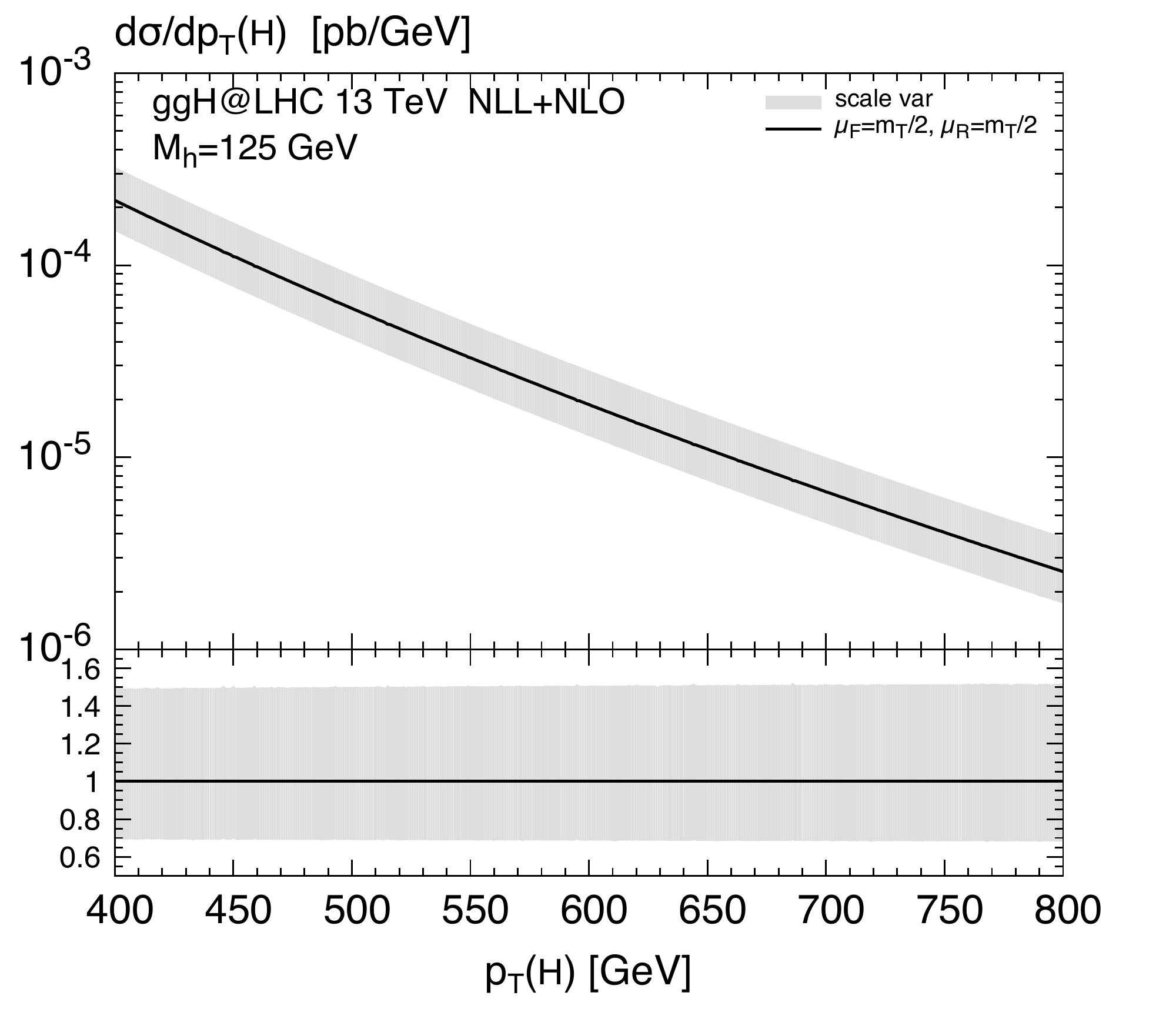} \\[-1em]
(a) & (b)
\end{tabular}
\caption[]{\label{fig:smscale}{SM prediction of the Higgs transverse momentum distribution at NLL+NLO for (a) $0$\,GeV$\le \pt \le400$\,GeV and (b) $400$\,GeV$\le \pt\le 800$\,GeV, with uncertainty bands due to scale variations as outlined in the text.}}
\end{center}
\end{figure}

In summary, our results for the $p_T$ spectrum depend on five scales:
the renormalization and factorization scales, and the three resummation
scales discussed above. In order to estimate the perturbative
uncertainty, we study the corresponding scale variations.  As far as
renormalization and factorization scales are concerned, the
uncertainties are estimated by performing the customary seven-point
$\muR$, $\muF$ variation, i.\,e. we consider independent variations
within the range $\mu_0/2 \le\muF,\muR\le 2\mu_0$ with
$1/2<\muR/\muF<2$. We then vary each of the three resummation scales
($Q_t$, $Q_b$, $Q_{\rm int}$) by a factor of two around their central values
(by keeping all the other scales to their central value). We finally
combine the ensuing four uncertainty bands by taking the envelope.
Figure \ref{fig:smscale} shows the reference SM prediction together with
its perturbative uncertainty. We see that the uncertainty ranges from
about $\pm 20\%$ at the peak, to about $+50\%$ $-30\%$ at $p_T=400$ GeV. 

\begin{figure}
\begin{center}
\begin{tabular}{cc}
\hspace*{-0.17cm}
\includegraphics[trim = 7mm -7mm 0mm 0mm, width=.36\textheight]{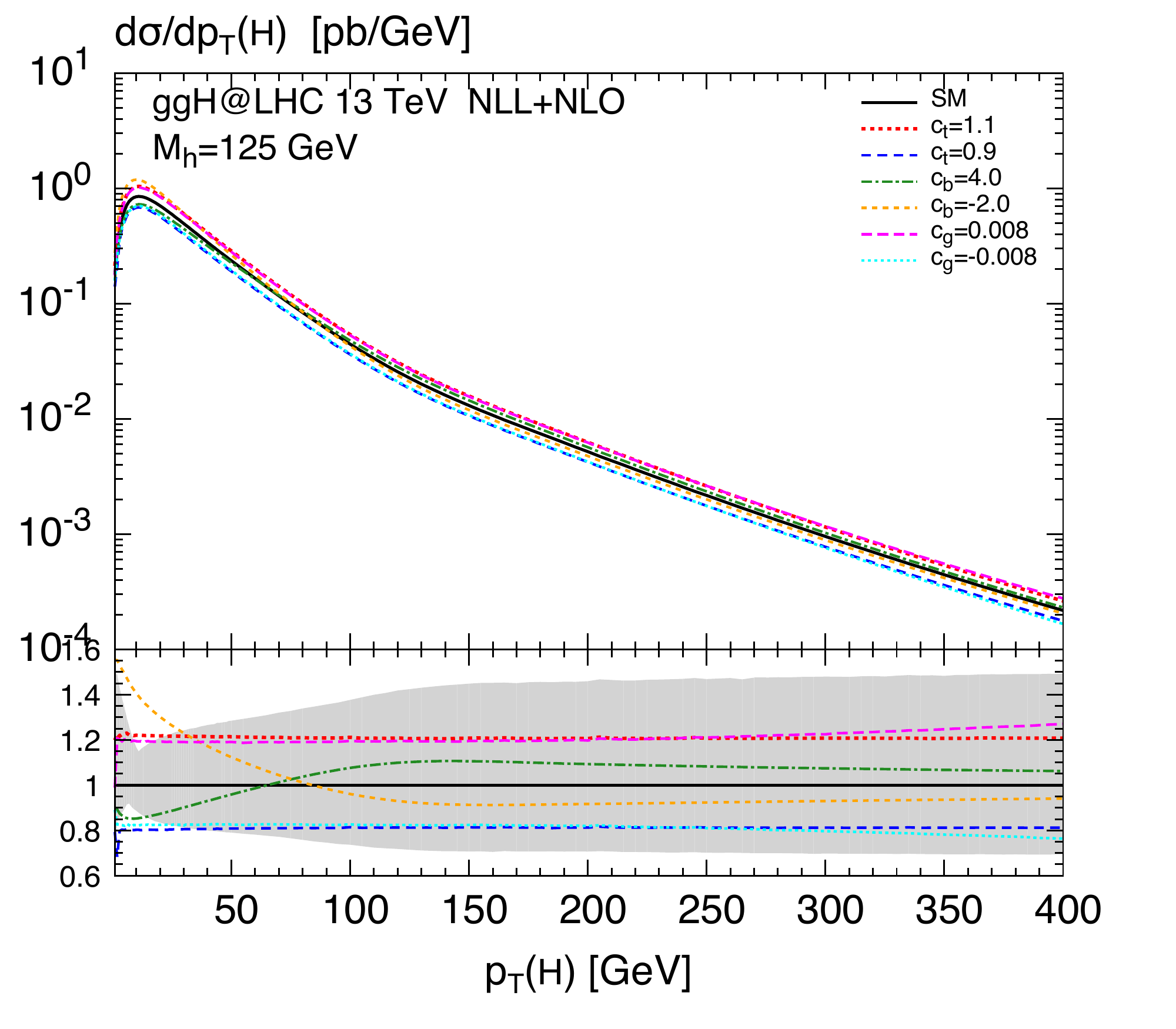} &
\includegraphics[trim = 7mm -7mm 0mm 0mm, width=.36\textheight]{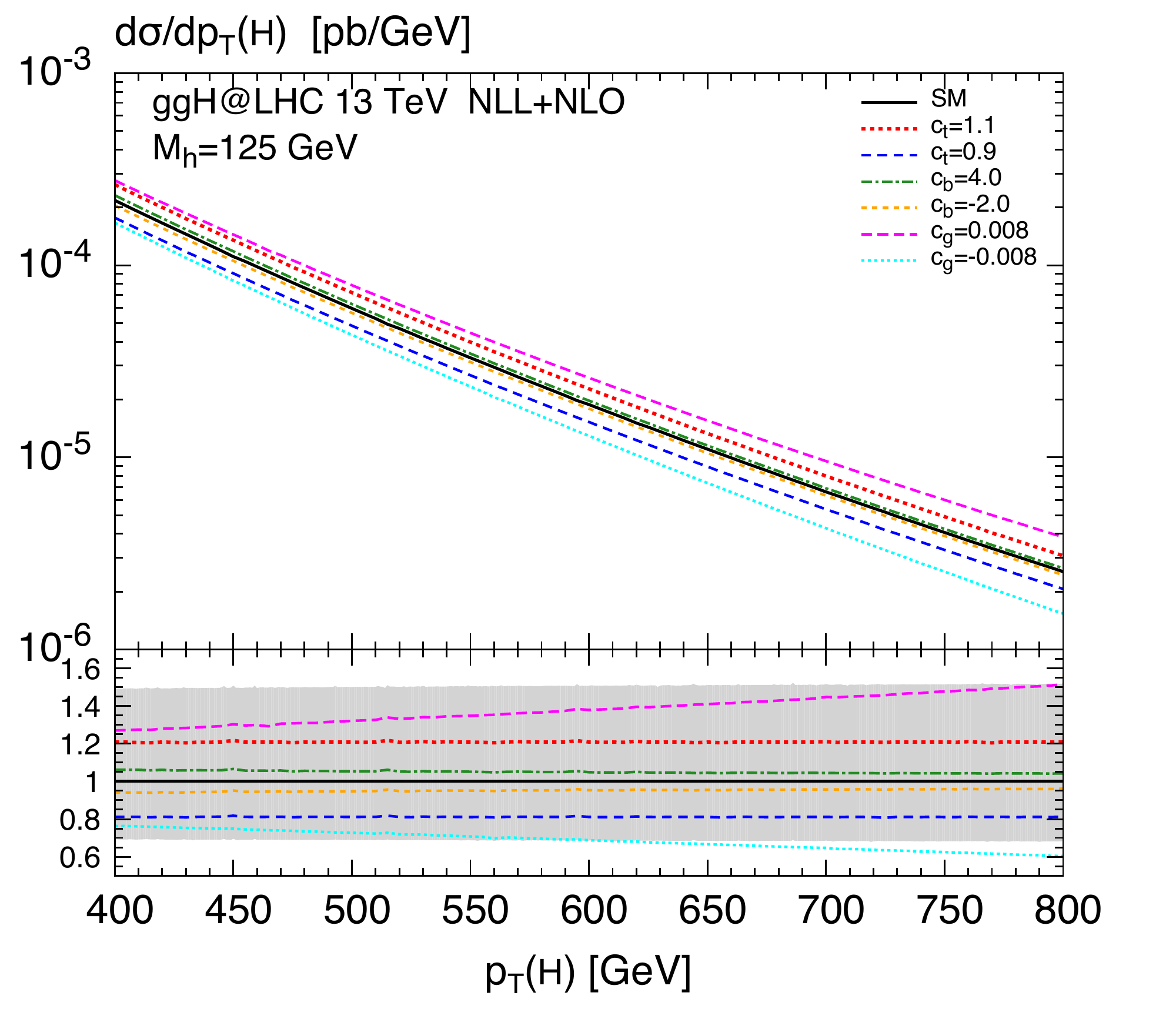} \\[-1em]
(a) & (b)
\end{tabular}
\caption[]{\label{fig:sep}{Higgs transverse-momentum spectrum in the SM 
(black, solid) compared to separate variations of the dimension-six operators 
for (a) $0$\,GeV$\le \pt \le400$\,GeV and (b) $400$\,GeV$\le \pt\le 800$\,GeV. 
The lower frame shows the ratio with respect to the SM prediction.
The shaded band in the ratio indicates the uncertainty due to scale variations. 
See text for more details.}}
\end{center}
\end{figure}

We start our analysis by considering the individual contribution of
exactly one operator. The values of the coefficients $c_t$, $c_g$ and
$c_b$ are varied as much as possible, while requiring the total cross
section (integrating over \pt{}) to not deviate by more than $20$\% from
the SM prediction, which is roughly the current experimental uncertainty
on the measured Higgs cross section. \fig{fig:sep} shows various
predictions of the Higgs transverse-momentum spectrum: SM (black,
solid), $c_t=1.1$ (red, dotted),  $c_t=0.9$ (blue, dashed), $c_b=4$
(green, dash-dotted), $c_b=-2$ (yellow, short-dashed), $c_g=0.008$
(magenta, long-dashed) and $c_g=-0.008$ (light-blue, short-dotted). The
lower frame illustrates the deviation from the SM prediction by taking
the ratio of the curves in the main frame to the black, solid one. The
grey-shaded band indicates the uncertainty of the SM result due to scale variations as
defined above.

Looking at the low-\pt{} interval ($0$\,GeV$\le \pt\le 400$\,GeV) in
\fig{fig:sep}\,(a) we can directly deduce from the  green, dash-dotted
and yellow, short-dashed curves that modifications of the bottom Yukawa
coupling through $c_b$ dominantly affect the low-\pt{} shape of the
distribution. In fact, at very low \pt{} we find effects that can even
exceed the uncertainty of the SM prediction. As expected, $c_b<1$ ($c_b>1$) 
softens (hardens) the spectrum in that region.\footnote{We point out, however, that this 
is true only when small deviations of $c_b$ from its SM value $c_b=1$ are considered. In this case the 
dominant effect of $c_b$ is on the top-bottom interference. When $c_b$ is significantly different from unity the squared bottom-loop contribution can change the picture.}
The point-like Higgs-gluon coupling $c_g$, on the other hand, modifies
the \pt{}-shape most notably at large transverse momenta ($400$\,GeV$\le
\pt\le 800$\,GeV), see \fig{fig:sep}\,(b), where a positive (negative)
$c_g$ value hardens (softens) the spectrum.  As expected, modifications
of solely the top Yukawa through $c_t$ have almost exclusively the
effect of a rescaling of the total cross section. 

By and large all the deviations from the SM prediction through the 
dimension-six operators are within the
scale uncertainty, although the differences in shape give some
additional sensitivity to distinguish such effects.
It is clear that in order to disentangle effects of this order
it is necessary to start from a more accurate
SM prediction. 

\begin{figure}
\begin{center}
\begin{tabular}{cc}
\hspace*{-0.17cm}
\includegraphics[trim = 7mm -7mm 0mm 0mm, width=.36\textheight]{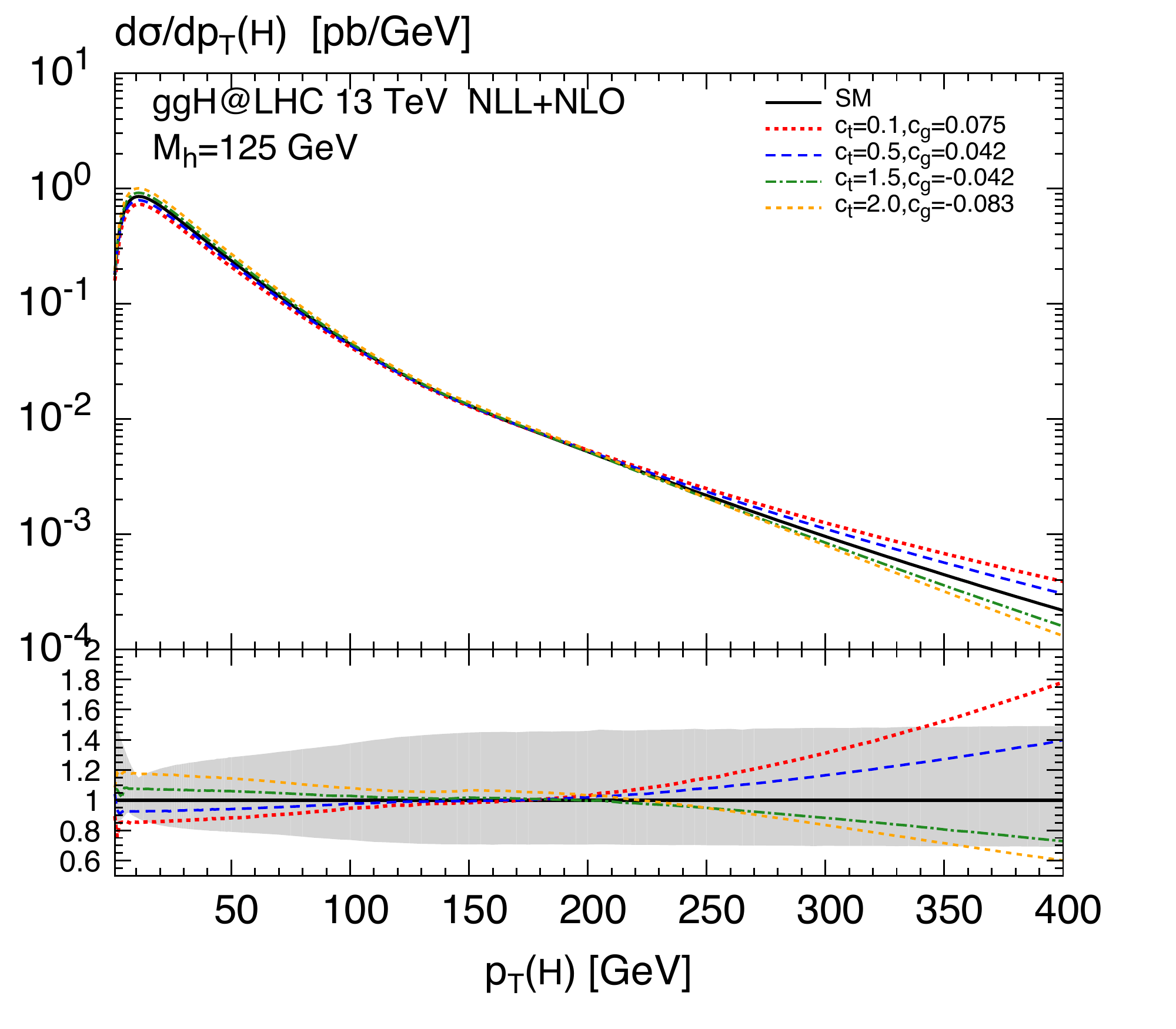} &
\includegraphics[trim = 7mm -7mm 0mm 0mm, width=.36\textheight]{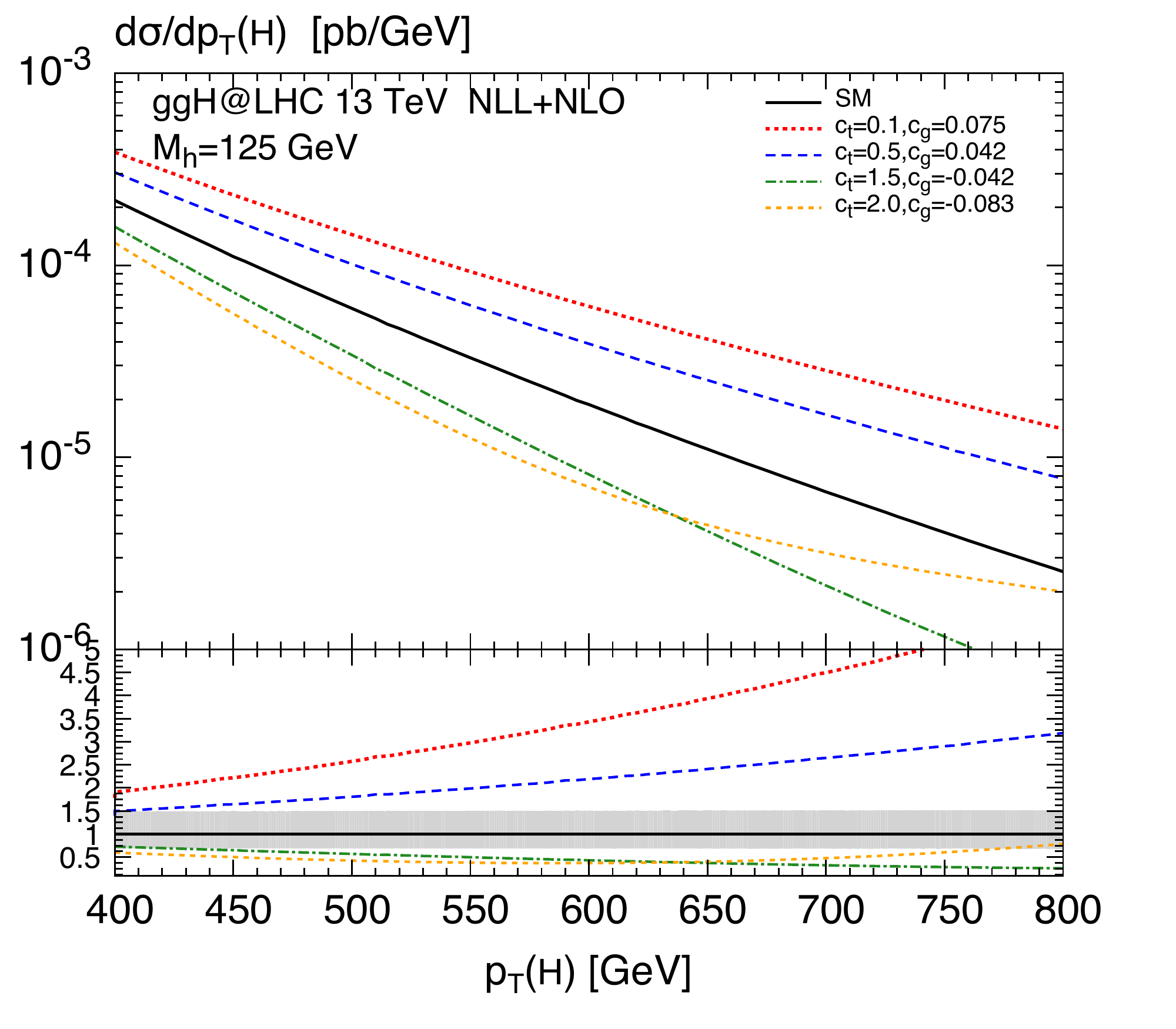} \\[-1em]
(a) & (b)
\end{tabular}
\caption[]{\label{fig:ctcg}{Higgs transverse-momentum spectrum in the SM 
(black, solid) compared to simultaneous variations of $c_t$ and $c_g$ 
for (a) $0$\,GeV$\le \pt \le400$\,GeV and (b) $400$\,GeV$\le \pt\le 800$\,GeV. 
The lower frame shows the ratio with respect to the SM prediction.
The shaded band in the ratio indicates the uncertainty due to scale variations. 
See text for more details.}}
\end{center}
\end{figure}

\begin{figure}
\begin{center}
\begin{tabular}{cc}
\hspace*{-0.17cm}
\includegraphics[trim = 7mm -7mm 0mm 0mm, width=.36\textheight]{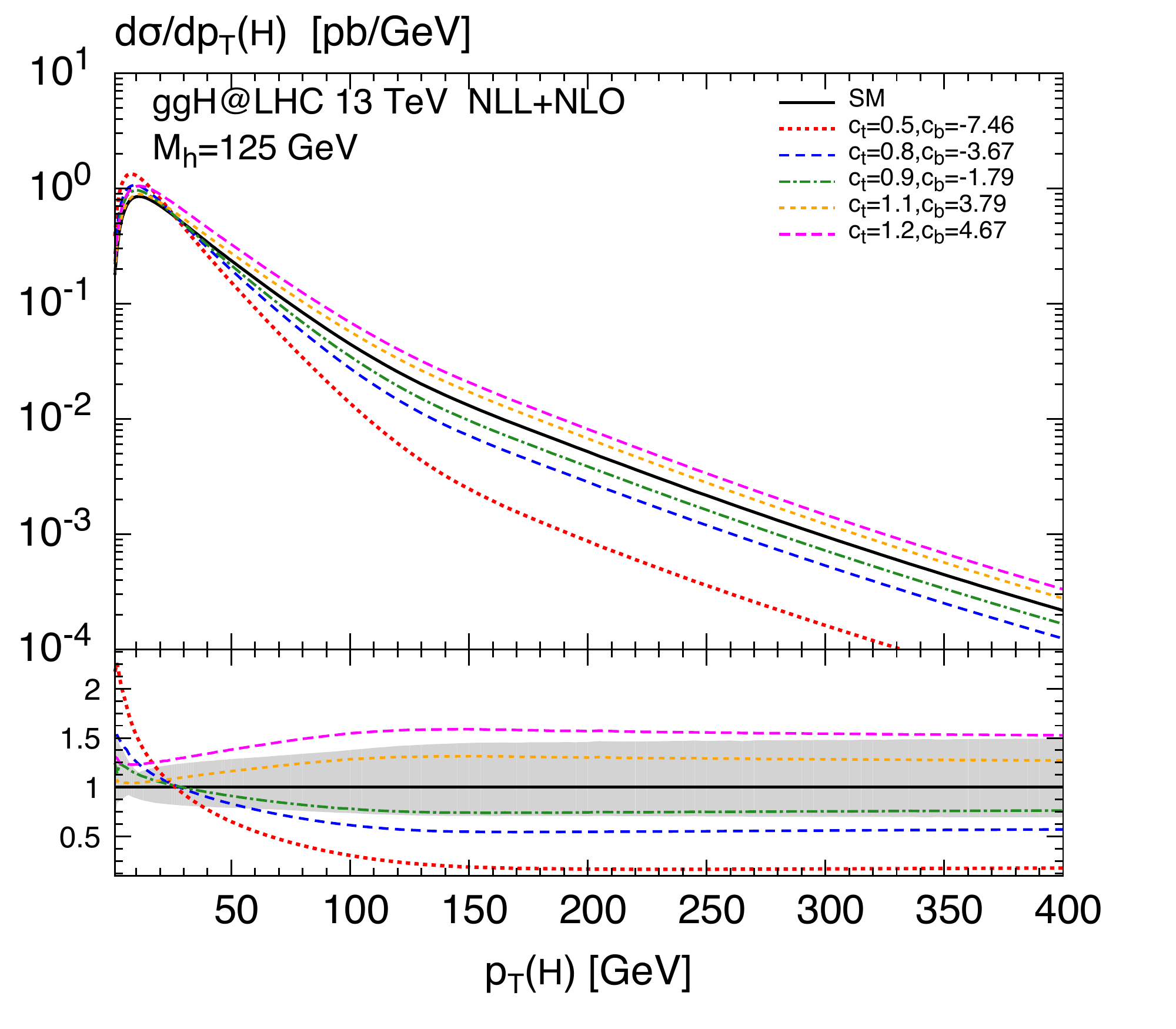} &
\includegraphics[trim = 7mm -7mm 0mm 0mm, width=.36\textheight]{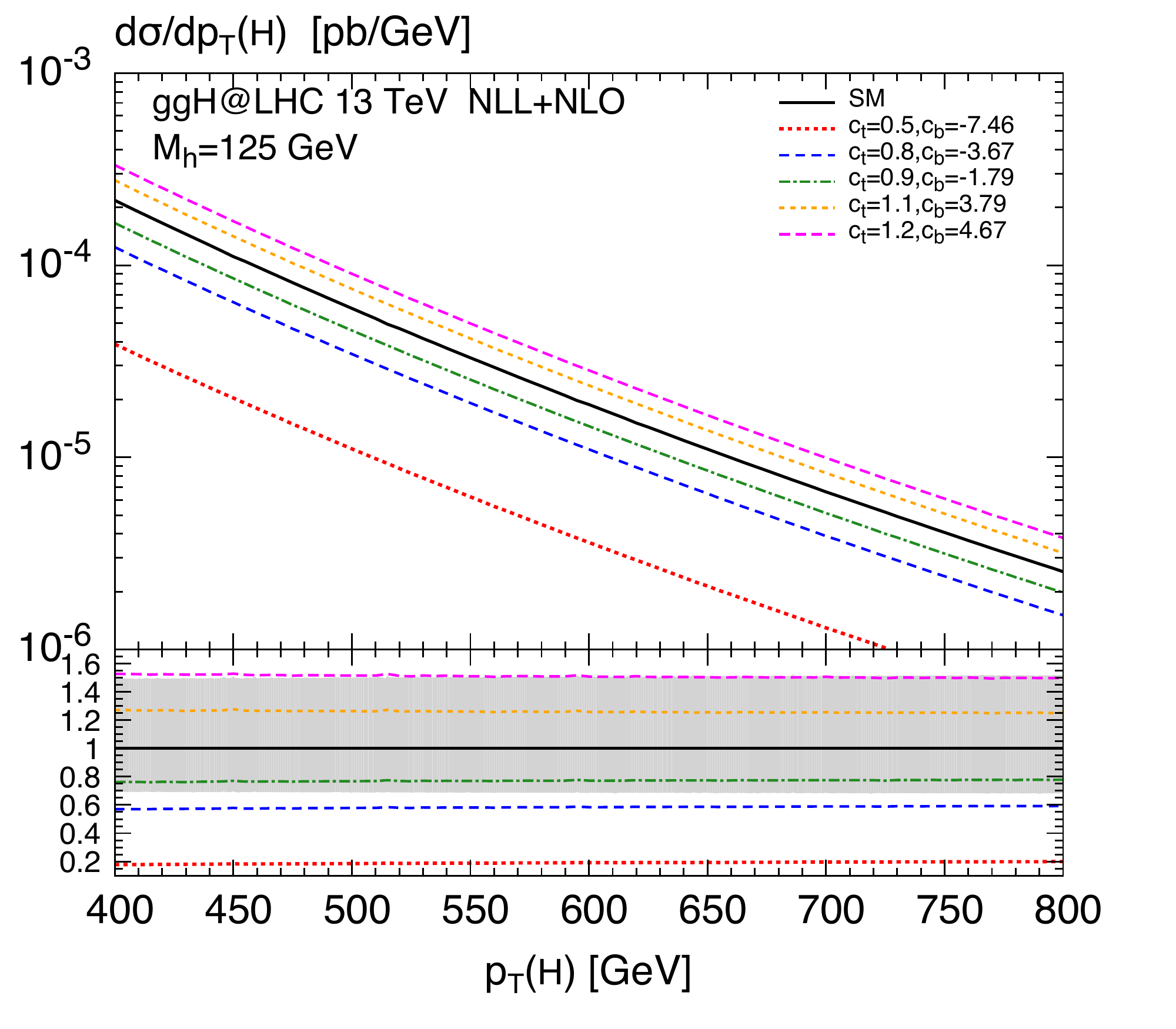} \\[-1em]
(a) & (b)
\end{tabular}
\caption[]{\label{fig:ctcb}{Higgs transverse-momentum spectrum in the SM 
(black, solid) compared to simultaneous variations of $c_t$ and $c_b$ 
for (a) $0$\,GeV$\le \pt \le400$\,GeV and (b) $400$\,GeV$\le \pt\le 800$\,GeV. 
The lower frame shows the ratio with respect to the SM prediction.
The shaded band in the ratio indicates the uncertainty due to scale variations. 
See text for more details.}}
\end{center}
\end{figure}

\begin{figure}
\begin{center}
\begin{tabular}{cc}
\hspace*{-0.17cm}
\includegraphics[trim = 7mm -7mm 0mm 0mm, width=.36\textheight]{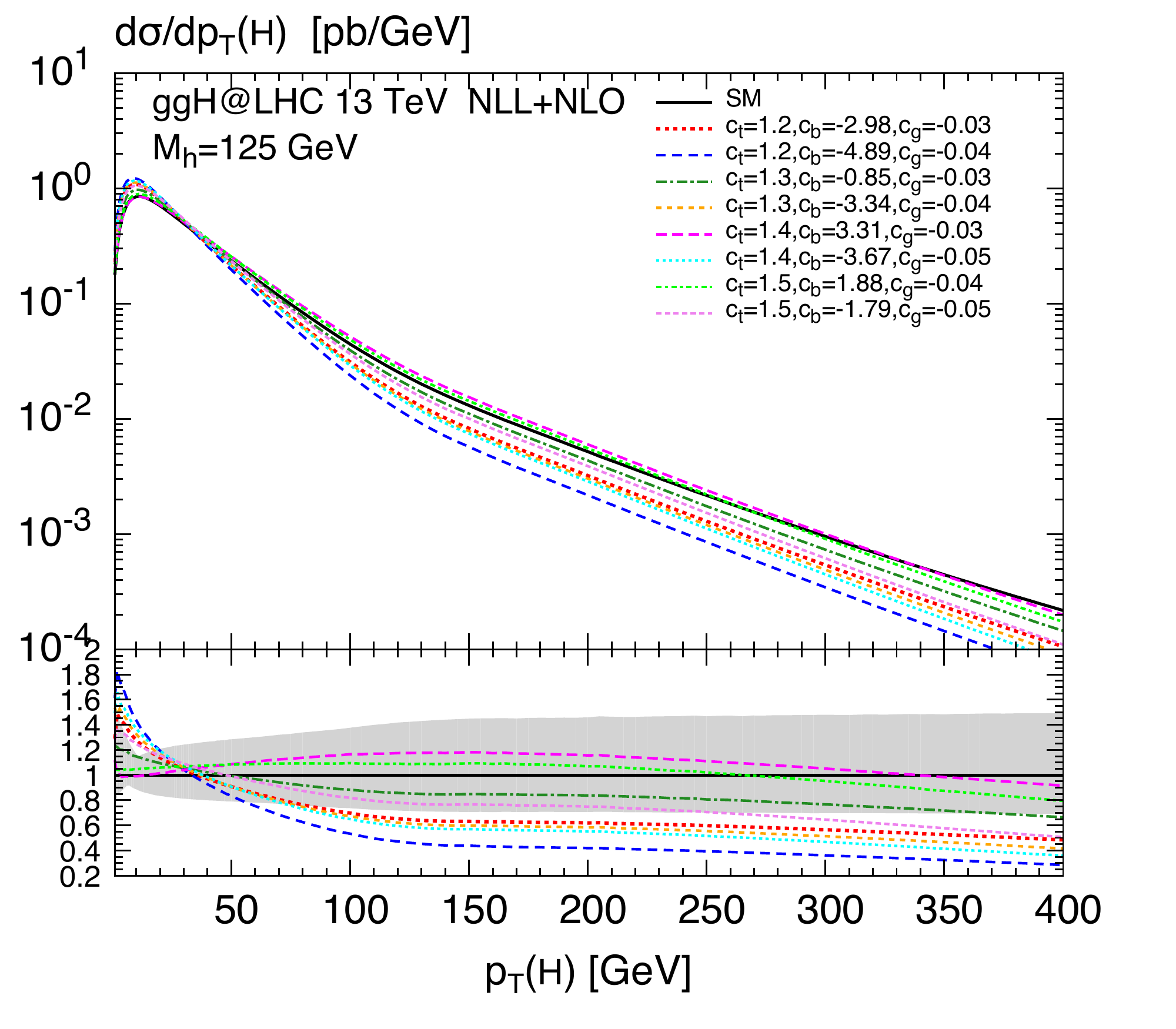} &
\includegraphics[trim = 7mm -7mm 0mm 0mm, width=.36\textheight]{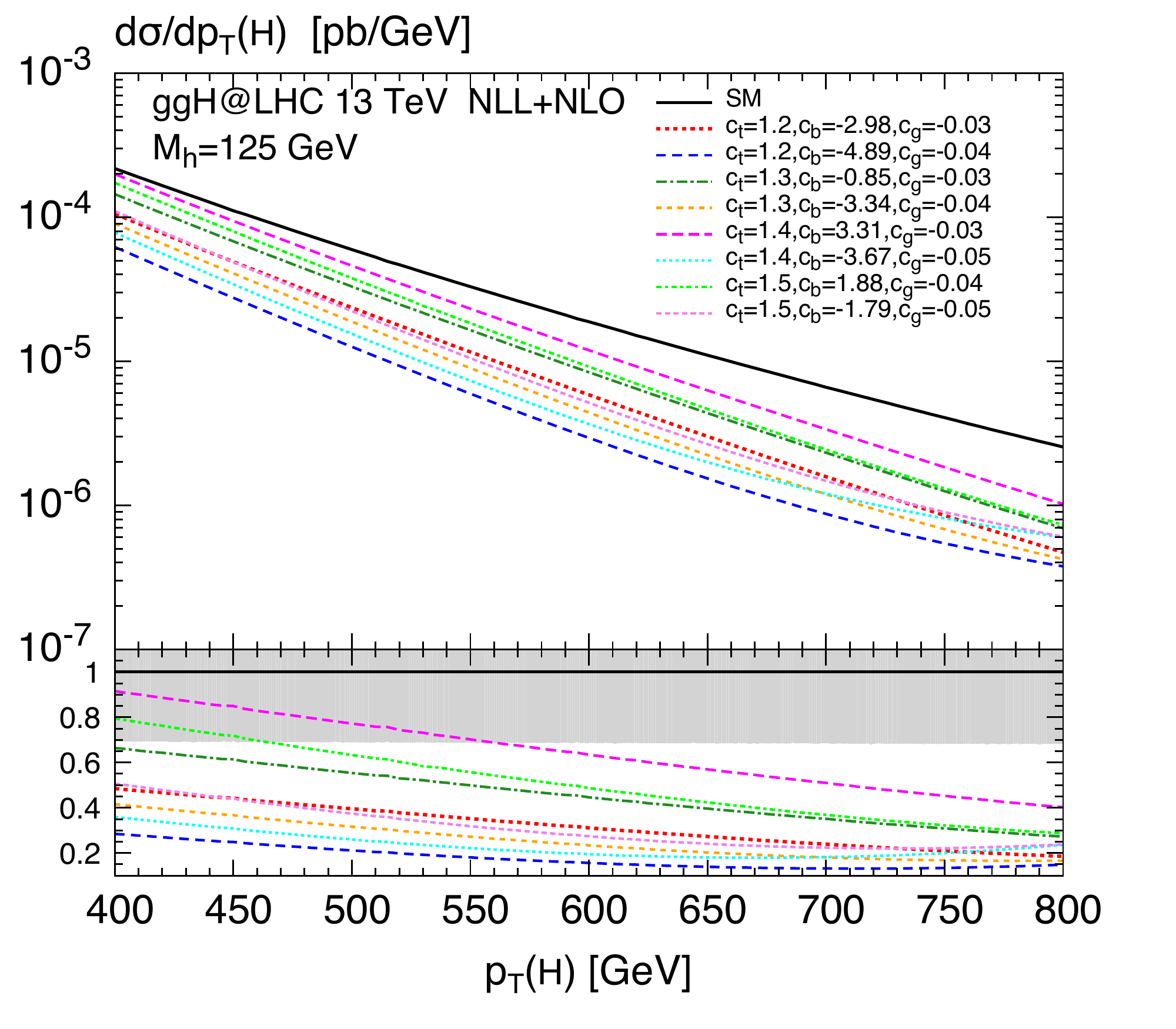} \\[-1em]
(a) & (b)
\end{tabular}
\caption[]{\label{fig:ctcgcb}{Higgs transverse-momentum spectrum in the SM 
(black, solid) compared to simultaneous variations of $c_t$, $c_g$ and $c_b$
for (a) $0$\,GeV$\le \pt \le400$\,GeV and (b) $400$\,GeV$\le \pt\le 800$\,GeV. 
The lower frame shows the ratio with respect to the SM prediction.
The shaded band in the ratio indicates the uncertainty due to scale variations. 
See text for more details.}}
\end{center}
\end{figure}

By contrast, the simultaneous variation of more than a single
coefficient, as considered in \figs{fig:ctcg}-\ref{fig:ctcgcb}, gives
rise to more significant effects.
The $c_g$, $c_t$ and $c_b$ parameters are chosen in the ballpark
suggested by the studies of Refs.~\cite{Dumont:2013wma,Falkowski:2015fla,Butter:2016cvz},
while still keeping the inclusive cross section within about 20\% of its SM value,
Indeed, many combinations of $c_g$, $c_t$ and $c_b$ can be found which
mildly affect the total cross section, while significantly changing
the shape of the spectrum. 

In \fig{fig:ctcg} we present the simultaneous variation of $c_t$ and
$c_g$. The general pattern of these figures follows the pattern of
\fig{fig:sep}, but for the variations: $c_t=0.1$, $c_g=0.075$ (red,
dotted); $c_t=0.5$, $c_g=0.042$ (blue, dashed); $c_t=1.5$, $c_g=-0.042$
(green, dash-dotted) and $c_t=2$ $c_g=-0.083$ (yellow, short-dashed). In
this case, both the small and high-$\pt$ behaviour of the spectrum is
altered by the different combinations of $c_t$ and $c_g$ coefficients.
It is clear that in particular the large-\pt{} region offers a
good discrimination between the different structures of $c_t$ and $c_g$ 
in terms of shape. Again, negative (positive) $c_g$ values
will soften (harden) the spectrum.  The effects are well beyond the
theoretical uncertainties already at NLL+NLO.  We note that the yellow,
short-dashed curve corresponding to $c_t=2$, $c_g=-0.083$ develops a
minimum in the ratio to the SM around $\sim 650$\,GeV. This is due to a compensation between
the negative interference between the ${\cal O}_1$ and ${\cal O}_2$
operators, which is proportional to $c_g c_t$ and the contribution of
${\cal O}_1$ itself, which is proportional to $c_g^2$ and tends to
produce a harder spectrum with respect to the SM prediction.

\fig{fig:ctcb} shows spectra with modified top and bottom
Yukawa couplings: $c_t=0.5$, $c_b=-7.46$ (red,
dotted); $c_t=0.8$, $c_b=-3.67$ (blue, dashed); $c_t=0.9$, $c_b=-1.79$
(green, dash-dotted); $c_t=1.1$ $c_b=3.79$ (yellow, short-dashed)
and $c_t=1.2$, $c_b=4.67$ (magenta, long-dashed). 
In this case, the compensation of the BSM contributions
is less straightforward, and it is difficult to compensate $c_t>1$
without significantly affecting the inclusive cross section.
For the magenta, long-dashed curve ($c_t=1.2$, $c_b=4.67$) we, thus, allow for a 
bigger change of the total cross section up to 30\%.
 As pointed out before, 
the bottom-loop softens the spectrum and, since the variation of the 
bottom Yukawa coupling is rather large, the squared bottom-term is larger 
than the top-bottom interference term and the spectrum is softened 
also when negative $c_b$ values are considered. The shape difference 
to the SM is very significant, but only in the small-$\pt{}$ region, where 
the soft-gluon resummation is crucial to obtain a reliable prediction. Indeed, the contribution
of the bottom loop decreases with growing $\pt{}$
\cite{Langenegger:2006wu} and above 150 GeV 
the spectra have all very similar shapes, $c_t$ driving their normalization.

Finally, we discuss spectra obtained by switching on all three SMEFT operators,
as shown in Fig.~\ref{fig:ctcgcb}: $c_t=1.2$, $c_b=-2.98$, $c_g=-0.03$ (red, dotted),  $c_t=1.2$, $c_b=-4.89$, $c_g=-0.04$ (blue, dashed), $c_t=1.3$, $c_b=-0.385$, $c_g=-0.03$
(green, dash-dotted), $c_t=1.3$, $c_b=-3.34$, $c_g=-0.04$ (yellow, short-dashed), $c_t=1.4$, $c_b=3.31$, $c_g=-0.03$
(magenta, long-dashed); $c_t=1.4$, $c_b=-3.67$, $c_g=-0.05$ (light-blue, short-dotted); 
$c_t=1.5$, $c_b=1.88$, $c_g=-0.04$ (light-green, short-dash-dotted) and 
$c_t=1.5$, $c_b=-1.79$, $c_g=-0.05$ (violet, very-short-dashed).
Our focus here is on scenarios with increased top-quark Yukawa coupling (up to $c_t=1.5$). 
These scenarios would be of particular interest in the case in which the excess on the $t\bar{t}H$
rate over the SM prediction \cite{ATLAS:2016awy,CMS:2016vqb} should be confirmed. 
In order to compensate the
increase in the cross section driven by $c_t>1$ a negative $c_g$ has been
chosen. We observe a general tendency of the BSM spectra to fall below the
SM prediction in the intermediate and high transverse-momentum regions,
which is due to the negative $c_g$ contribution. The total rate is
compensated by the enhancement in the low $\pt{}$ region, due to a combination 
of the negative $c_g$ coefficient with both negative and positive $c_b$ modifications.
Overall, we find sizable distortions of the $\pt{}$ shapes due to the dimension-six 
operators far beyond the scale uncertainties of the NLL+NLO SM prediction, 
that exceed the previously considered scenarios with two simultaneous varied coefficients 
in both size and significance. Despite the similar overall behavior, the predictions for 
the various scenarios may differ significantly, which enables their discrimination when 
compared to data.

We conclude this Section with a comment on the validity of the EFT approach. The computation we have performed is carried out under the assumption that we can consider the effects of higher-dimensional operators as a ``small'' perturbation with respect to the SM result. This implies in particular that
the effect of dimension-eight operators can be neglected. This is not obvious, given that
we are studying also the large transverse-momentum region.
To check the above assumption we have repeated our calculations by dropping the
${\cal O}(1/\Lambda^4)$ suppressed terms originating from
the square of the dimension-six contributions. We find that in most of the cases
the differences with respect to the results shown in Figs.~\ref{fig:sep}-\ref{fig:ctcgcb} are very small, even at high transverse momenta. Only in the scenarios considered in Fig.~\ref{fig:ctcg} ($c_t=0.1, c_g=0.075$ and $c_t=2, c_g=-0.083$) the ${\cal O}(1/\Lambda^4)$ effects are important, and thus, the corresponding
quantitative results should be interpreted with care.

\section{Summary}
New Physics might be not accessible at the LHC through direct searches, e.g., with
the discovery of new resonances. In that case, it is crucial to fully exploit the data
to study possible (small) deviations from the SM predictions.
SMEFT offers a formalism
for the parametrization of high-scale BSM effects, which can be used for this purpose.
In the SMEFT framework BSM effects
are parametrized through appropriate higher-dimensional operators, and bounds on the corresponding Wilson coefficients can be set by comparing to the experimental data.

In this paper, we have presented a computation of the transverse-momentum 
spectrum of the Higgs boson in which the SM prediction is supplemented by possible BSM effects.
Such effects are modeled by augmenting the 
SM Lagrangian with appropriate dimension-six operators. Our calculation consistently 
includes all the terms up to ${\cal O}(\as^3)$ accuracy and is supplemented by 
soft-gluon resummation at NLL accuracy, which is required to obtain reliable predictions at small transverse momenta.  At the same level of accuracy we implement three
dimension-six operators, related to the modifications of top and bottom Yukawa couplings and to the inclusion of 
a point-like $ggH$ coupling. Additionally, we studied the impact of 
the chromomagnetic operator on the Higgs cross section at LO, which had been previously addressed in the literature
by different groups with contradicting results.

We have performed a comprehensive study of the possible effects due to the different
dimension-six operators, by studying the impact of variations of $c_t$, $c_b$ and $c_g$ 
on the transverse-momentum spectrum of the Higgs boson. We varied the above coefficients in
the range suggested by recent global analyses 
and required the total cross section 
to meet the SM prediction at NLO within the current ${\cal O}(20\%)$ experimental 
uncertainty. Our results can be summarized as follows:
\begin{itemize}
\item Variations of different SMEFT operators  
manifest themselves in different regions of the Higgs \pt{} spectrum. A modification of the Higgs-bottom 
Yukawa coupling (${\calO}_3$) induces sizable effects almost exclusively at small transverse momenta. 
A direct coupling of the Higgs boson to gluons (${\calO}_1$), on the other hand, changes the shape of the 
distribution most notably in the high-$p_T$ tail. As expected, changes in the top-quark Yukawa coupling 
(${\calO}_2$) primarily affect the normalization and approximately correspond to a simple rescaling of the spectrum.
\item The shape of the transverse momentum distribution depends on the mass of the particle that mediates the Higgs-gluon 
coupling. The lower the mass of that particle, the softer is the resulting spectrum. Therefore, the 
$\pt{}$ shape associated with the bottom loop is softer, in particular at small transverse momenta, than the SM one and,
when increasing the absolute value of the bottom-quark Yukawa coupling, positive (negative) values soften (harden) 
the spectrum, if the top-bottom interference is dominant (small variations of $c_b$). In contrast the spectrum becomes always 
softer for $|c_b|\gg 1$, if the squared bottom-loop is dominant (large variations of $c_b$). 
Furthermore, a point-like coupling between gluons and the Higgs boson leads to the hardest spectrum and 
a positive (negative) $c_g$ value hardens (softens) the shape as compared to a Higgs boson mediated by a top-quark loop.
\item While individual variations of the various Wilson coefficients produce
effects that hardly exceed the NLL+NLO perturbative uncertainties, the simultaneous variation 
of two or more operators can significantly distort the spectrum, still keeping
the total rate consistent with the NLO prediction within the current experimental uncertainties.
\item Choosing combinations of $c_t$, $c_b$ and $c_g$ that compensate each other at the level of the total
cross section allows us to deform the shape of the Higgs \pt{} spectrum far beyond 
the uncertainties of our NLL+NLO prediction in the SM.
\end{itemize}

We conclude our discussion by adding a few comments on the limitations of
the calculation presented here.
When only small deviations from the SM are considered the theoretical uncertainty
affecting the SM prediction becomes relevant. We have seen that at NLL+NLO the uncertainties from missing higher-order contributions, estimated through scale variations, are about $\pm 20\%$ at the peak and increase by roughly a factor of two at high transverse momenta.
The natural question is whether
the calculation we have carried out can be extended to the next order, i.e., to NNLL+NNLO.
To consistently carry out such extension
we would need the heavy-quark mass effects at NNLO, which are currently unavailable. A possible way out is to include the effects beyond NLL+NLO in the HTL, as is
currently done in state of the art SM calculations \cite{Grazzini:2013mca}.
This approach implies that the relevant Higgs production amplitudes would 
contain a $c_g$ term already in the SM.
Nonetheless, it is reasonable to assume that
the QCD effects beyond our NLL+NLO accuracy factorise with respect to the BSM corrections.
In this approximation, the relative BSM/SM effects we have obtained in this paper (i.e., the ratios plotted in the lower panels of Figs.~\ref{fig:sep}-\ref{fig:ctcgcb}) can be directly used to include BSM effects on top of NNLL+NNLO accurate SM predictions.

Another aspect which deserves some comments is the set of dimension-six operators we have considered.
In the present calculation we have limited ourselves to consider the contributions of the operators related to modified top and bottom Yukawa couplings and of the additional direct $Hgg$ interaction. As discussed in Sec.~2, although formally suppressed by two powers of the top Yukawa coupling, the chromomagnetic operator could still significantly contribute, within the current bounds, to the gluon fusion cross section. 
The extension of our calculation to include these effects is left for future work.

\noindent {\bf Acknowledgements.} We would like to thank Gino Isidori and Fabio Maltoni for useful discussions and comments on the manuscript. This work is supported by the 7th Framework Programme of the European
Commission through the Initial Training Network HiggsTools
PITN-GA-2012-316704.

\bibliographystyle{UTPstyle}
\bibliography{pT_EFT}

\end{document}